\documentstyle[12pt]{article}
\topmargin - 2.5cm
\begin{document}

\begin{center}
Exact solutions for classical few-body systems from 
the multichannel  quantum inverse problem \\
 Zakhariev B.N.,  Chabanov V.M.  \\
Laboratory of the
Theoretical Physics, Joint Institute for Nuclear Research  Dubna, 141980,
Russia;\\
e-mail: zakharev@thsun1.jinr.ru; chabanov@thsun1.jinr.ru; homepage:
http://thsun1.jinr.ru/~zakharev  
\end{center}

 \begin{abstract}
{\small
 A surprising "duality" of the Newton equation with {\it time-dependent}
forces  and the {\it stationary} Schr\"odinger equation is discussed. 
Wide classes of exact solutions not known
before for few-body Newton equations are generated  directly
from exactly solvable
multichannel models discovered in quantum mechanics due to the inverse
problem and the supersymmetry (SUSYQ) approach. 
The application of this duality to the control of the stability 
(bifurcations) of classical motions is suggested.   }
 \end{abstract}

\section{Introduction}

In this paper the results related to the control of classical few-body
systems are presented. We use the duality of corresponding solutions with
the known exact solutions of quantum problems. The complete sets of 
such exactly solvable models have been recently intensively discussed 
\cite{Z,ChZ}. The transition between quantum and classical solutions is 
possible through simply  renaming variables and functions.

Physics is permanently developing in a tight interplay with mathematics
("the laws of nature are written in the mathematical language"). There 
are cases when the same mathematical equation can describe physical systems 
of different nature, which broadens and deepens their understanding. We 
shall demonstrate this for the example of the multichannel formalism that is 
a powerful and universal tool in quantum inverse and direct problems of 
few-body physics \cite{Z,ChZ,ZhZ}.

We start with briefly reminding  the one-dimensional and one-channel case.

The corresponding stationary Schr\"odinger equation
\begin{eqnarray}
-\psi ''(x) + v(x)\psi (x) = E \psi (x)
\label{Schr}
\end{eqnarray}
can be rewritten by changing the notation $$\psi \to z;\,\,\, x\to t;\,\,\,
d^{2}/dx^{2}\to d^{2}/dt^{2};\,\,\, [v(x)-E]\psi \to F(t,z)$$ so that
it will look like an equation of motion of a classical particle under the
action of the force $F(t,z)$ dependent on time $t$ and coordinate $z$:
\begin{eqnarray}
  \ddot z(t) = [v(t)-E] \enskip z \equiv F(t,z).
\label{Newt}
\end{eqnarray}
It is a Newton equation: the acceleration of a particle moving along the
 z axis is determined by the force $F(t,z)$ linearly dependent on z
(an oscillator-type potential $\sim \, z^{2}$) with a time-dependent 
strength.
In a classical  equation, the former quantum energy $E$ and the 
potential $v(x)$ become, respectively, a parameter and a function
characterizing the external force.

This information was obtained at the School on Quantum Mechanics in 
Mexico, 1998, from Prof. Bogdan Mielnik and Dr. David Fernandez and 
 paper \cite{Mieln}, see also references therein.
It was surprising for us and particularly interesting because it means that
complete sets of exactly solvable quantum models \cite{ChZ,ChZ2}
simultaneously provide exact solutions to classical problems of quite 
a different nature.

It is also a "paradox" that the eigenvalues $E_{n}$ for {\it stable} bound
states in quantum mechanics correspond to the points of bifurcations 
(absolute instability) of
solutions for classical motion. Let us comment on this statement.
  At an energy slightly lower than a ground state energy level $E_{n}$, 
the nonphysical solution of the Schr\"odinger equation decreasing 
asymptotically to the left grows exponentially to the right.
 And somewhat  above the energy level, the solution with the 
same asymptotic behavior on the left bends more strongly, acquires an
additional node and grows exponentially to the right with another sign.
In the classical case it corresponds to an instantaneous change of the 
regime for a particle motion
when the parameter $E$ goes through the point $E_{n}$ : a particle
being initially at the origin in the first case goes to the right with
 increasing accelerations  and in the second case it goes to the left.
 At the point $E=E_{n}$ the particle localized previously  at the
origin after the influence of the time-depending external
field returns to the origin.

So, the possibility of shifting, creating and removing arbitrary energy levels 
in quantum systems ~\cite{Z,ChZ,ChZ2}
corresponds in the classical case to the ability to control the bifurcation
points which are the parameters of principal importance characterizing the 
instability of motion. 

Of course, at the same value $E$, a classical scattering with
another initial ($t \rightarrow -\infty $) condition is possible 
corresponding to another dual quantum asymptotic ($x \rightarrow -\infty $)
condition.

In the quantum inverse problem, the potential is uniquely determined by
a complete set of spectral parameters : energy level values, normalizing
parameters (spectral weights) and  phase shifts (or resonance positions
if S-matrix has fractional-rational form \cite{ZS}) for scattering.
Therefore,  classical counterparts of these parameters (e.g. bifurcations)
are control levers of the corresponding classical system. This interesting
fact was not so evident from the point of view of Newton equations.

 The intuition developed in exactly solvable quantum models \cite{Z,ChZ2}
allows one to qualitatively predict  also classical motion  without
computation.

\section{Multichannel, multi-dimensional and few-body systems}

 The multichannel systems \cite{F,ZS,B,A,KS} for vector-valued wave
functions with partial channel components $\psi_{i}(x)$  are simply 
matrix generalizations of the one-dimensional Schr\"odinger equation
(\ref{Schr})
\begin{eqnarray}
-\psi ''_{i}(x) + \sum_{j} v_{ij}(x)\psi_{j} (x) = E_{i} \psi_{i} (x),  
\nonumber \\
E_{i} \equiv E-\epsilon_{i}  
\label{system}
\end{eqnarray}
where the interaction matrix $v_{ij}(x)$ replaces the ordinary potential
and $\epsilon_{i}$ are threshold energies that can be  different in
partial channels.
This system can be
rewritten as multi-dimensional or few-body equations of classical mechanics
with special forces.

Let us replace the partial channel wave
function $\psi_{i}(x)$ and its space coordinate variable $x$ in the system
(3) by the coordinate $z_{i}$ of the $i$th classical particle and time:
$z_{i}(t)$.
Then the second derivative $\psi_{i}(x)$ is substituted by the acceleration
$\ddot z_{i}$ and all other terms  can be considered as the forces
$F_{i}(t,z)$ dependent on time and space coordinates acting on the $i$th
particle. So, elements of the potential matrix  $v_{ij}(x)$  and 
channel energy terms $E_{i}\psi_{i}(x)$ become
 constituents of the forces
$$F_{ij}(t) =\sum_{j} v_{ij}(t)z_{j}(t) - E_{i}z_{i}(t).$$

The whole system (3) becomes a system of equations
for several classical particles, where
 the particle accelerations are determined by functions of the 
time-dependent forces
  \begin{eqnarray}
 \ddot z  _{i}= \sum_{j} v_{ij}(t)z_{j}(t)-E_{i}z_{i}(t)
\equiv F_{i}(t).
\label{clsys}
 \end{eqnarray}
Again as in (\ref{Newt}) the parameters $E_{i}$ are no longer the channel
energies, but simply enter as parameters into the definition of the
forces acting in the system.  Therefore, every exact solution of the
multichannel quantum problem (and  complete sets of them) 
corresponds to an exact solution of the few-body classical problem
with time-dependent forces. {\it The asymptotic and boundary conditions of 
quantum problems determine initial and final conditions of the
corresponding classical solutions}.  For the
classical scattering solutions it is possible to use nonphysical quantum
solutions growing {\it linearly} in asymptotic regions: $\psi(x)=a x + b$ (for
 $E_{i}=0$).  The constants $a$ and $b$ fixing the inclination of the 
straight line and the position point of the node of the wave function 
determine in the classical problem the position and velocity of a 
particle in some moment of time (for instance, t=0): $z \, = \, at \, + \, b$.

As an example of the prediction of a peculiar behavior 
of few-body systems, one  can choose the dual model of multichannel exact 
solution when one of the partial channel wave functions is concentrated 
at the origin. This corresponds to increasing the spectral weight 
parameter of this partial channel 
$C_{i} \equiv \frac{d \psi _{i}(x)}{dx}|_{x=0}$.
We have discovered \cite{ChZ2} that in this case all other
partial components of wave functions are transferred to the chosen $i$th
channel. In the limiting case $C_{i} \to \infty$ the whole wave function 
is totally gathered in the $i$th channel, so that all other channels are
"emptied". In the classical interpretation, it means that it is easy to 
determine the forces $F_{i}(t)$ which move mainly only one chosen 
$i$th particle.  Although these forces are strong enough to expect that 
all other particles are not passive spectators of intensive motion of 
their chosen companion.

To give an idea of the character of explicit formulae of
exactly solvable multichannel models arising from the inverse problem,
we show an example of the reflectionless interaction matrix with one
bound state at energy $E_{bound}<\epsilon_{i}$:

\begin{eqnarray}
 v_{ij}(x)=2\frac{d}{dx}\frac{M_{i} M_{j}
e^{-(\kappa_{i}+\kappa_{j}) x}}
{1+\sum_{m}\frac{M_{m}^{2}}{2 \kappa_{m}} e^{- 2 \kappa_{m} x}}, \\
\kappa_{i}=\sqrt{\epsilon_{i}-E_{bound}},  \nonumber
\label{vij}
\end{eqnarray}
where $M_{i}$ stand for   partial channel spectral parameters which are
pre-exponential factors in the decreasing asymptotic tails of
the bound state wave functions. An interested reader can find
other formulae for a complete set of inverse problem and SUSYQ
transformations in \cite{ChZ}.

 The transformation of the three-channel system of type (3)
by renaming: $$\psi_{1}(x) \rightarrow \,x(t), \enskip  
\psi_{2} \rightarrow  \, y(t), \enskip 
\psi_{3} \rightarrow \, z(t)$$ gives us
the classical equations for the three-dimensional motion of a particle.

As interesting instructive examples can be considered the classical 
counterparts of quantum
systems with "paradoxical" coexistence of bound and scattering states at
the same energy \cite{ChZ2}, 1999 or combinations of
absolute transparency and strong reflection for different 
linearly independent solutions of multichannel equations.

 It is possible to significantly extend the class of exactly
solvable models if we introduce, into the quantum equations 
nonhomogeneous terms which can be treated as sources with an
arbitrary dependence on coordinates.
From the classical point of view, they correspond to forces
$F_{n}(t)$ which are independent of coordinates. Having the sets of independent
exact solutions of the homogeneous equations we get the expressions for Green
functions. They give explicit analytic  expressions for
the solutions of the initial nonhomogeneous equations.

The quantum multichannel systems obtained by adiabatic expansions
have derivatives of first order which can be interpreted as friction
forces depending on the velocity.  But  exact solutions of
these systems remain an open problem yet.

\section{Conclusion}

The authors consider it a pleasure and  an honor to participate in this
issue of the Few-Body Systems dedicated to Prof. W.~Gl\"ockle, the 
scientist having record
results in this fundamental field of quantum physics. Already his first work
\cite{Gl} produced a strong impression on us.  It was a significant
contribution to the multichannel formalism. Prof. W.~Gl\"ockle also
stimulated our investigations on qualitative theory of quantum design 
reviewed later in the books "Lessons on Quantum Intuition" , \footnote{
With the subtitles:  Gallery of wonderful
potentials. Algorithms of spectral, scattering and decay control. Exactly
solvable models of inverse problems and SUSYQ. Universal building elements
"bricks and blocks" of quantum systems.}
"New ABC of Quantum Mechanics (in pictures)" \footnote{With the subtitles: 
Qualitative
foundations of the wave literacy. Construction of quantum systems with
the given properties.} and in
a series of review articles \cite{Z,ChZ}. We appreciate 
very much that Prof. W.Gl\"ockle was one of the first physicists
in the Western Europe who organized, at Ruhr University at his seminar
 consideration of our theory.

\section{Acknowledgments}

The authors are thankful to Prof. Bogdan Mielnik and Dr. David Fernandez
for the information about the previous works on classical interpretation
of the quantum equations and to Prof. Man'ko for discussions of
these problems. The work of Chabanov V.M. was supported in part by
the INTAS grant 96-0457 within the research program of the International
Center for Fundamental Physics in Moscow.

\end{document}